\documentclass[12pt]{article}\usepackage{amsmath,amsfonts}
\usepackage{amsmath}
\usepackage{amssymb}
\usepackage{amsfonts}
\usepackage{eucal}

\newcommand{\be}{\begin{equation}}
\newcommand{\bse}{\begin{subequations}}
\newcommand{\ese}{\end{subequations}}
\newcommand{\bea}{\begin{eqnarray}}
\newcommand{\eea}{\end{eqnarray}}
\newcommand{\ba}{\begin{array}}
\newcommand{\ea}{\end{array}}
\newcommand{\ee}{\end{equation}}

\makeatletter \@addtoreset{equation}{section}

\makeatother
\def\lambdab{\lambda^{\!\!\!\!-}}
\def\ie{{\it i.e. }}
\def\eg{{\it e.g. }}

\begin{document}
\baselineskip 18pt%

\begin{titlepage}
\vspace*{1mm}%
\hfill%
\vbox{
    \halign{#\hfil        \cr
             hep-th/0605110 \cr
             IPM/P-2006/046 \cr
           }
      }
\begin{center}
{\Large{\bf Inherent Holography In Fuzzy Spaces}}\\ {\bf and}\\
{{\large\bf An $N$-tropic Approach To The  Cosmological Constant Problem}}%
\vspace*{10mm}%

{\bf \large {M. M. Sheikh-Jabbari}}%

\vspace*{0.4cm}
{\it {Institute for Studies in Theoretical Physics and Mathematics (IPM)\\
P.O.Box 19395-5531, Tehran, IRAN}}\\
{E-mail: {\tt  jabbari@theory.ipm.ac.ir}}%
\end{center}

\begin{center}{\bf Abstract}\end{center}
\begin{quote}
In this letter we show that  the noncommutative spaces, and in
particular fuzzy spheres, are natural candidates which explicitly
exhibit the holography, by noting that the smallest physically
accessible  volume is much larger that the expected Planckian
size. Moreover, we show that fuzzy spheres provide us with a new
approach, an ``$N$-tropic'' approach, to
 the cosmological constant problem, though in a Euclidean
 space-time.
\end{quote}


\date{\today}
\end{titlepage}
\textwidth 16.5cm%
\section{Introduction}

Despite the extensive work devoted to and partial progress in some
areas, the problems arising when gravity and the quantum theory
should both be employed has escaped a thorough understanding. The
two areas of interest in this direction are understanding the
blackhole Bekenstein-Hawking (BH) entropy and the cosmological
constant problem. In the former case  string theory (the leading
candidate  of quantum gravity) has provided us with a nice (but
partial) resolution of the problem \cite{BHentropy}.

On the other hand, based on the BH entropy formula it has been
argued that in a theory of gravity the number of physical degrees
of freedom should scale with the area around any region of the
space, rather than the volume which is what we see in an ordinary,
non-gravitational field theory. This idea has been named {\it the
holographic principle} \cite{holography}. (For a review on more
recent developments on holography in general, and holographic
bound in particular, see \cite{Bousso}.)  Although the idea of
holography came through blackhole analysis, it has been promoted
to the guiding principle for finding or distinguishing correct
formulation of quantum gravity. According to the holographic
principle, any theory of quantum gravity should exhibit
holography.

Attempts in constructing holographic quantum gravity, except the
cases which are related to string theory has not been successful.
Within string theory, however, the AdS/CFT duality \cite{AdS/CFT}
and its variants are the best known examples in which holography
is realized \cite{SW}.

 In this note we put forward the idea that
the holography and the fact that the number of physical degrees of
freedom is growing with the area of the space surrounding any part
of space, rather than its volume, has something to do with the
quantum nature of space-time itself. This can be realized in a
class of {\it noncommutative} space-times. In other words, due the
noncommutativity of the space, there is a smallest  cell in our
space (which can only carry one bit of information) and hence
information can't be squeezed further. This as we will see, in
part, leads to a ``geometric'' realization of the holographic
principle. That is, in our setting holography is connected with
the inherent nature of space-time and in a sense has a kinematical
appearance rather than a dynamical one.

As the other very interesting outcome of our noncommutative
setting we show that the noncommutative fuzzy spheres appear as
vacuum solutions to a Euclidean gravity Matrix theory \cite{Nair}.
In this theory, and in the ``continuum'' limit, the Ricci
curvature of the spherical vacuum solution in Planck units becomes
the Cosmological Constant (CC). As in the fuzzy spheres (some
power of the) radius is quantized in Planck units, the
cosmological constant in our model is
 integer-valued and hence is stable against (continuous)
quantum corrections. This provides a solution to the ``technical
naturalness'' of the CC problem, \ie how the CC is stabilized
under quantum corrections.

This article is organized as follows. In the next section we
present a definition of fuzzy spheres and show how the holography
appears as an inherent property of them. We also discuss an
interesting noncommutative flat space limit of these fuzzy
spheres. We then turn to the cosmological constant problem and
present a gravity theory in which fuzzy spheres appear as vacuum
solutions. We show within this theory the {\it Euclidean} CC
problem has found an answer.

\section{A kinematical realization of holography}

The idea we would like to demonstrate here is that the
noncommutative spaces, and in fact a subclass of them, the fuzzy
spheres, exhibit the interesting property of accommodating much
less degrees of freedom than one would expect from their
commutative counterpart.

To see these we need a simple definition of fuzzy spheres. This
was given in Appendices B and D of \cite{TGMT}. It is based on the
fact that geometric $S^d$ is equivalent to the quotient
$SO(d+1)/SO(d)$. And the $S^d_F$ is the {\it quantized},
``fuzzified'' or ``discretized'' version of the $d$-sphere in such
a way that the $SO(d+1)$ invariance remains intact. This can be
achieved noting the fact that $SO(d+1)$ is a compact group and has
finite dimensional unitary representations. To put this idea at
work we note that an ordinary $d$ sphere can be defined by its
embedding coordinates in
a $d+1$ flat space as \cite{TGMT}%
\be\label{com.def}%
\begin{split}
\sum_{i=1}^{d+1} x_i^2&=R^2\\
\{x_{i_1},x_{i_2},\cdots, x_{i_d}\}&=R^{d-1}\epsilon_{i_1i_2\cdots
i_{d+1}} x^{i_{d+1}}
\end{split}
\ee%
where the bracket is the Nambu $d$-bracket defined among $d$
functions $F_i$ on a $d$ dimensional manifold with coordinates
$\sigma_j$, $F_i=F_i(\sigma_j)$, as
\be
\{F_1,F_2,\cdots, F_d\}=\epsilon_{i_1i_2\cdots i_d}
\partial_{i_1}F_1\partial_{i_2}F_2\cdots \partial_{i_d}F_d.
\ee%
As it is seen for $d=2$ it simply reduces to a Poisson
bracket. Note that in  \eqref{com.def} we have used the only two
invariant tensors of $SO(d+1)$, namely $\delta_{ij}$ and
$\epsilon_{i_1\cdots i_d}$.

The fuzzy spheres are then defined by turning the embedding
coordinates $x_i$'s into operators, which can be represented by
$N\times N$ matrices, and quantizing the Nambu bracket (this
parallels the steps of going from classical to quantum mechanics
when we replace Poisson brackets with the commutators). The
quantized Nambu brackets (QNB), for even $d$ are defined as
totally anti-symmetrized products of $d$ operators or matrices
(for odd $d$ the situation is more involved but does have
a solution \cite{TGMT}), \ie%
\be%
 [F_1,F_2,\cdots, F_d]\equiv \epsilon_{i_1i_2\cdots i_d}
F_{i_1}F_{i_2}\cdots F_{i_d}.%
 \ee%

In moving from Poisson bracket to commutators we introduce a
constant of nature which is $\hbar$. Here when we fuzzify a given
sphere, \ie when we move from Nambu brackets to QNBs, we need to
introduce an ``$\hbar$'', which we will denote by $\lambdab$.
Explicitly \cite{TGMT},%
\be%
 \{F_1,F_2,\cdots,F_d\}\stackrel{Quantization}{-----\rightarrow}
\left(\frac{1}{i\lambdab}\right)^{d/2}[F_1,F_2,\cdots, F_d]%
\ee%
In our set up we would like to think $\lambdab$ as a constant of
nature, similarly to the $\hbar$. $\hbar$ is a measure of
quantization in the phase space, whereas $\lambdab$ is a measure
of quantization on quantum space-time. One may use $\lambdab$ and
$R$ to define the ``fuzziness'' $\ell$ which is a short distance
length scale (one may
think of it as the Planck length):%
\be\label{lambda-def}%
\lambdab=\frac{\ell}{R}%
\ee%
In sum, an $S_F^d$ is defined through $N\times N$
hermitian matrices $X_i,\  i=1,2\cdots, d+1$ such that%
\begin{subequations}\label{NC.def}%
\begin{align}
\sum_{i=1}^{d+1} X_i^2&=R^2 {\bf 1}_{N\times N}\\
[X_{i_1},X_{i_2},\cdots, X_{i_d}]&={L}^{d-1}\epsilon_{i_1i_2\cdots
i_{d+1}} X^{i_{d+1}}.
\end{align}
\end{subequations}%
Note that ${L}$ is different than $\ell$. Using
Eqs.\eqref{com.def}, \eqref{lambda-def} and \eqref{NC.def} it is
seen that%
\be\label{L-ell-R}%
L^{d-1}=\ell^{d-1}\ \left(\frac{R}{\ell}\right)^{\frac{d-2}{2}}\ .%
\ee%
(For the fuzzy odd spheres the Nabmu bracket technology can still
be used, however  here we'll restrict ourselves to the even $d$
cases, see \cite{TGMT, Torab} for more details).

 One may try to solve eqs.\eqref{NC.def} to find explicit fuzzy
sphere solutions. This can be done using some group theory and in
particular representation theory for $SO$ groups, \eg see
\cite{Torab, SF-soln}. Intuitively, the effects of ``fuzziness''
means that there is a cut-off on the highest ($SO(d+1)$) angular
momentum on the $S^{d}_{F}$. If we call this highest angular
momentum by $n$, then one can show that \cite{Torab, SF-soln}
\be\label{largest-J}%
R\sim \ell n %
\ee%
for large $n$. The exact relation for even $d$ is \cite{SF-soln}
\be\label{R-n-exact}%
R^2=\ell^2 n(n+d).%
\ee%
Combining the above
with \eqref{lambda-def} we see that $\lambdab=1/n$; in other words
$\lambdab$ is inverse of the largest angular momentum possible on
the $S^d_F$.

Moreover, the group theory analysis  shows that \emph{size of the
matrices} $N$ is growing like $n^{d-1}$ for large
enough $n$ ($R\gg \ell$), that is, for any $d$%
\be\label{R-l-relation}%
R^{d-1}=\ell^{d-1} N .%
\ee%
As we see, as a result of the fuzziness, radius of the fuzzy
sphere is quantized in units of the fuzziness $\ell$. The
classical continuum sphere is then recovered in the $\ell\to 0$
keeping $R$ fixed. In this limit $N$ goes to infinity and hence
$\lambdab$ also vanishes.

Equation \eqref{NC.def} bears an interesting statement of
holography. This stems from the fact that in any field theory
defined on the fuzzy sphere we are dealing with operators in
$N\times N$ representation. In order to see the holography we
argue in what follows that the smallest volume that one can probe
on the noncommutative fuzzy sphere is not $\ell^d$ but a much
larger volume element $\mathcal{V}_{min}$ which is proportional to
$n^{-d/2}$ in units of the sphere volume.

To see how such a $\mathcal{V}_{min}$  arises in the fuzzy sphere
setup one may consider a specific flat space limit of the sphere,
under which the $S^d_F$ goes over to a kind of {\it Lorentz
invariant} noncommutative $d$-plane, $\mathbb{R}^d_{\lambdab}$.
Explicitly,
consider%
\be\label{limit}%
\ell\to 0, \ R\to\infty,\ \ L^{d-1}R\equiv {\mathcal{V}}_{min}={{\rm} fixed}.%
\ee%
The claim is that $\mathcal{V}_{min}$ is the smallest volume one
can probe on the sphere. This can  be checked recalling
\eqref{NC.def}. In the ``intermediate'' scales, \ie $ \ell\ll
X_i\ll R$,  one can always expand the $X$'s  about a North pole,
{\it e.g.} $X_{d+1}\sim R$ and $X_i\ll R,\ i=1,2,\cdots, d$,
explicitly:%
\[X_{d+1} = n\ell {\bf{1}}+\frac{1}{R} \delta X_{d+1},
\]
 The above constitutes the
generalization of the stereographic projection of the two sphere
to the fuzzy $d$-spheres. The above ``$d$ dimensional fuzzy
stereographic projection'' for the case of fuzzy two sphere has
been discussed in \cite{Sphere-Moyal}. It is now straightforward
to check that in the above mentioned limit one can  relax
(\ref{NC.def}a),
and (\ref{NC.def}b) now reads as%
\be\label{NC-d-plane}%
[X_{i_1},X_{i_2},\cdots,X_{i_d}]= L^{d-1}R\ \epsilon_{i_1i_2\cdots
i_{d}}\ {\bf 1}.
\ee%

From the above, noting the definition of the QNB and with a little
bit of  algebra (which is very similar to the one leading to
Hiesenberg uncertainty relation starting from
$[x,p]=i\hbar$),  one can show that%
\be\label{lowest-volume}%
\Delta V_d\equiv \Delta X_1\Delta X_2\cdots \Delta X_d\geq L^{d-1}R=\mathcal{V}_{min}, %
\ee%
\ie  $\mathcal{V}_{min}$ is the smallest volume on the
$\mathbb{R}^d_{\lambdab}$. As the expansion about the North pole
is quite generic, one can therefore conclude that
$\mathcal{V}_{min}$ is also the smallest volume on the $S^d_F$.
\footnote{Another way to see that $\mathcal{V}_{min}$ is the
smallest volume is to recall \eqref{R-n-exact} and
(\ref{NC.def}a). If one of the $X$'s, say $X^{d+1}$ is taking its
maximal value $\ell n$, the others should all be of the order
$(\ell n)^{1/2}\sim (\ell R)^{1/2}$. Using some basic facts about
the $SO$ groups representation theory this means that $\Delta
X_i\sim \sqrt{\ell R}$ and hence $\Delta V_d\geq (\ell
R)^{d/2}=L^{d-1}R$ where in the last equality \eqref{L-ell-R} has
been used.} Equation \eqref{limit} can then be
written as%
\be\label{R-l-L}
\frac{Vol(S^d_F)}{\mathcal{V}_{min}}=\left(\frac{R}{L}\right)^{d-1}%
\ee%
where the left-hand-side is the number of smallest cells one can
fit into the $S^d_F$ of volume $R^d$.

Equation \eqref{R-l-L} is indeed a statement of holography. To see
this let us consider a $d+2$ dimensional Schwarzchild blackhole
with ADM mass $M$. The horizon of this
blackhole is a $d$ dimensional sphere with radius $R_{H}$:%
\be\label{ADM-mass}%
l_P M= \left(\frac{R_{H}}{l_P}\right)^{d-1}%
\ee%
The Bekenstein-Hawking entropy, i.e. the area of the horizon in
Planck units is then%
\be%
S_{B.H.}\sim \left(\frac{R_{H}}{l_P}\right)^{d}%
\ee%
 If the {\it horizon is an $S^d_F$} of
radius $R_H$ with $\ell=l_P$, using \eqref{L-ell-R} and
\eqref{R-l-L}, one can readily check that%
\be%
S_{B.H.}=(\#{\rm\ cells\ on\ the\ }S^d_F)^2\ .%
\ee%
The above is basically the statement of holography. (Note also
that using \eqref{ADM-mass} and \eqref{R-l-relation} $l_P M\sim N$
size of the matrices describing the fuzzy sphere horizon.)

It is also worth noting that the $X_i$'s and any function of them
should be treated as quantum operators, which admit $N\times N$
representation and hence any theory on the $S^d_F$ is a matrix
theory. Moreover, noting that $\mathcal{V}_{min}=L^{d-1} R=(\ell
R)^{d/2}$ involves both the UV character $l_P$ and the IR
character $R$ one expects the IR/UV mixing phenomenon to show up
in the
formulation of any theory on this background \cite{progress}.

\section{ An $N$-tropic Approach to the Euclidean Cosmological Constant problem}

In this section we present a Matrix {\it Euclidean} gravity theory
which has the fuzzy spheres among its vacuum solutions. Since this
theory has been discussed in some detail in \cite{Nair} we only
sketch the ideas of this theory and present its action.

This gravity theory is based on the Mansouri-Cheng ``gravity as
gauge theory''\cite{F-Mansouri}: To obtain an ordinary gravity
theory which has a group manifold $G/H$ as its vacuum solutions,
we start with coordinates $x_i,\ i=1,2,\cdots, d\equiv dim G-dim
H$ which are in the (infinite dimensional unitary) representation
of the Lie algebra of $G$, here will be denoted by $g$. The
gravitational degrees of freedom are encoded in the vierbein
$e_i^a(x),\ a=1,2,\cdots, d$ and the connection
$\Omega_i^\alpha(x)$, $\alpha=1,2,\cdots, dim H$ (in general the
$\alpha$ index is running from one to $dim\ {\rm Env} G-d$, where
${\rm Env} G$ is the enveloping algebra for the fundamental
representation of $g$). They appear through the covariant
derivative ${\cal D}_i$ \be {\cal D}_i=\partial_i+e_i^a(x) T^a+
\Omega_i^\alpha(x) I^\alpha \ee%
 where $I^\alpha\in h$ form a
complete basis for the fundamental representation of the Lie
algebra of $H$, $h$, and  $T^a$ the basis for $g-h$, hence $(T^a,
I^\alpha)$ form a complete basis for $g$. (In more general cases
the set of $I^\alpha$ should be extended, so that $(T^a,
I^\alpha)$ covers the Enveloping algebra for the fundamental
representation of $g$, ${\rm Env} G$.) As such they are $d\times
d$ unitary matrices. The gravity action is then constructed from
gauge invariant powers (of commutators) of ${\cal D}_i$. In our
example $G=SO(5)$ and $H=SO(4)$ and since the enveloping algebra
of $G$ is other than $G$, it is $U(4)$, $a$ is running from one to
four and $\alpha$ from one to 12. In our example the most natural
from for the action is the Chern-Simons gravity \cite{Nair}.

Within the above approach formulation of gravity on a
noncommutative ``fuzzified'' geometry is straightforward, {\it if
$G$ and $H$ are compact groups}. In order that it is only enough
to recall that for such groups it is always possible to find
finite dimensional unitary $N\times N$ representation which is
naturally embedded in $u(N)$. Hence the coordinates $x_i$,
$e_i^a(x)$ and $\Omega_i^\alpha(x)$ are all turned into $N\times
N$ unitary matrices which are in general  non-commuting. ${\cal
D}_i$ are then taking values in $U(d)\otimes U(N)$. ($U(d)$ is the
enveloping algebra of $G$.) The curvature two-form in the
non-commuting case can again be defined as ${\cal F}_{ij}=[{\cal
D}_i, {\cal D}_j]$. As the $x_i$'s, and hence the derivatives
$\partial_i$, are non-commuting, ${\cal F}_{ij}$ has a constant
piece. That is this part which leads to the cosmological constant
term in the gravity action.  ${\cal F}_{ij}$ has also a piece
which is proportional to $I^\alpha$. This part contains the
Riemann curvature two form ${\cal R}_{ij}$ and a part proportional
to $T^a$ which is the torsion \cite{Nair}.

In our case, where $d=4$ and $G/H=SO(5)/SO(4)$, we choose $
I^\alpha=\{i\gamma^5,\gamma^a\gamma^5,\gamma^{ab}, i{\bf 1}\}$,
$T^a=i\gamma^a$ and
for the action we take the Chern-Simons action%
 \be\label{action}
S=\kappa \frac{1}{N}{\rm Tr}(\gamma^5\epsilon_{ijkl}{\cal F}_{ij}{\cal F}_{kl})%
\ee%
where the ${\rm Tr}$ is over both the $4\times 4$ and  $N\times N$
matrices. After expanding the above action in terms of the Riemann
curvature and the torsion, what we find is an Einstein-Hilbert
gravity action plus a cosmological constant and some torsional
terms (see eq.(42) of \cite{Nair}). The torsion terms are
proportional to the fuzziness and hence go away in the continuum
limit. The demand that in the continuum (large $N$) limit, and
after proper scaling of the gauge fields and coordinates, we
should recover the usual Einstein gravity, upon assumption
$\ell=l_P$, fixes $\kappa$ as $\kappa^{-1}=R^2 \ell^2$ which is
equal to the cosmological constant. The vacuum solutions to the
above gravity theory, by construction, include the fuzzy four
sphere whose volume and the Cosmological Constant $\Lambda$, are
related as  $\Lambda^{-1}= R^4N^{-2/3}=\mathcal{V}_{min} $.
Therefore, the cosmological constant whose value is tied to the
number of degrees of freedom (or the size of the matrices) is
fixed and being quantized is protected against perturbative
quantum corrections. In this sense the cosmological constant, as
the size of matrices, is a constant of nature like $l_P$.

One should, however, note that quantization of the CC (in  Planck
units) in itself is not enough to solve the CC problem. For
example, within the string theory setup of flux compactifications
\cite{Bousso-Polch} the value of the four dimensional CC is
proportional to the fluxes and hence quantized. But, in that case,
unlike ours, there are extremely large number of possibilities
which leads to the string theory ``landscape'' \cite{landscape}.
The CC problem then re-appears as how/why one of these
possibilities is realized.

Within our approach, however, it is not explicitly seen how the
gravity theory given by \eqref{action} manages to overcome the
usual problem about the contribution of the tadpole diagrams and
zero point energies to the CC. The answer should definitely lie in
the fact that in our gravity theory both UV and IR dynamics of the
gravity are modified due to the noncommutativity which in part
forces us to add some other terms, \eg torsion, to the Einstein
gravity. Exploring this line is postponed to future works.

\section{Discussion}

In this short note we have tried to convey some ideas regarding
holography which in part leads to a new approach to the CC
problem. First we discussed holography in our setup. If we assume
that the horizon of a $d+2$ dimensional Schwarzchild
blackhole is an $S^d_F$ with $l_P=\ell$, we find that\\
\centerline{ $S_{B.H.}\sim$ \emph{Area of horizon in Planck
units=(Area of horizon in units of ${\mathcal{V}}_{min})^2$}.}

As for the CC problem,  we showed if we assume that the
``quantum'' version of the {\it Euclidean} de Sitter space is a
noncommutative fuzzy sphere, for the four dimensional case, the CC
in the Planck units is equal to ${\lambdab}^{-2}$. In our setup,
if we choose $\ell=l_P,\ R=$Hubble radius today, then
$\mathcal{V}_{min}=(submilimeter)^{4}$, where $\mathcal{V}_{min}$
is the smallest {\it observable} volume, and $\lambdab=10^{-60}$.
Let us now suppose we have the energy density $\rho$. It
gravitates as   $G_{N}\rho=l_P^2\rho$. Equating this to (the
square root of) the smallest observable volume
$\mathcal{V}_{min}$, one obtains that $\rho\sim (few\times
TeV)^4$. This is essentially the same argument that is behind the
ADD large extra dimensions proposal \cite{ADD}.  This is very
interesting, because within our model, similarly to the ADD large
extra dimension models, one can hope to see observable effects
from the quantum structure of the space-time already at the level
of LHC, and much lower than the Planck energy. Such effects are
basically arising from corrections \eg to the Standard Model
written on the noncommutative $\mathbb{R}^4_{\lambdab}$. These
corrections should appear through dimension six operators to the
Standard Model and are suppressed by powers of $(l_P^3 R)^{1/2}$.
(This could be seen from the fact that a field theory on
$\mathbb{R}^4_{\lambdab}$ has essentially the form of a field
theory on noncommutative Moyal plane, for a review on the latter
\eg see \cite{Szabo}.) Working out the details of this point and
the phenomenological implications of our model to physics at LHC
is a very interesting direction to be explored in future works.

One may also wonder whether the (geometric) picture of the
holography we presented here, which is based on fuzzy spaces
(spheres), can be reconciled with the best formulated example of
holography, \ie the AdS/CFT? There are some pieces of evidence
that there are indeed some, yet uncovered, connections between the
two, \eg the fact that the radius of the sphere in the
$AdS_p\times S^q$ spaces for $(p,q)=(5,5), (4,7)$ and $(7,4)$ in
Planck units is the same as what we have for fuzzy spheres, \ie
$R^{q-1}=l_P^{q-1} N$ \cite{AdS/CFT}. This hint is suggesting that
in the eventual picture for ``quantum space-times'' which is
emerging via dual gauge theories the spheres in the AdS spaces
turn into fuzzy spheres. There are some partial evidence in this
regard for the $AdS_5\times S^5$ case arising from the Tiny
Graviton Matrix Theory \cite{TGMT, Torab}.

Finally,  we would like to stress that here we only discussed the
Euclidean case. The discussion on the cosmological constant does
not go through for the Minkowski signature, \ie the fuzzy de
Sitter space $dS^4_F$ case, as for this case we should take
$G/H=SO(4,1)/SO(3,1)$ and $SO(4,1)$ is non-compact and has no
finite dimensional unitary representation. The formulation
developed in \cite{Nair-2006} may, however, help in this
direction. Despite of that, the arguments about the holography and
in particular eq.\eqref{NC.def}, the $\mathbb{R}^4_{\lambdab}$
space and eqs.\eqref{limit},\eqref{lowest-volume} (with a bit more
care) may also be used for the Lorentzian signature.

\section*{Acknowledgements}

I have discussed many issues about the ideas written up here
during the last four years with many people especially Savas
Dimopoulos, Robert Brandenberger and Simeon Hellerman. I would
like to thank them for encouraging me to write them up. I would
like to thank Esmaeil Mosaffa, Parameswaran Nair for helpful
comments and especially Eva Silverstein  for comments and a
suggestion for the title.


\end{document}